# Foundation of an analytical proton beamlet model for inclusion in a general proton dose calculation system

W. Ulmer[1,2] and B. Schaffner[1,3]

[1]Varian Medical Systems, Baden, Switzerland, [2]MPI of Biophysical Chemistry, Göttingen and Klinikum Frankfurt/Oder, Germany, [3]ETH Zürich

E-Mail: waldemar.ulmer@gmx.net

**Abstract:** We have developed a model for proton depth dose and lateral distributions based on Monte Carlo calculations (GEANT4) and an integration procedure of the Bethe-Bloch equation (BBE). The model accounts for the transport of primary and secondary protons, the creation of recoil protons and heavy recoil nuclei as well as lateral scattering of these contributions. The buildup, which is experimentally observed in higher energy depth dose curves, is modeled by inclusion of *two* different origins: 1. Secondary reaction protons with a contribution of ca. 65 % of the buildup (*for monoenergetic protons*). 2. Landau tails as well as Gaussian type of fluctuations for range straggling effects. All parameters of the model for initially monoenergetic proton beams have been obtained from Monte Carlo calculations or checked by them. Furthermore, there are a few parameters, which can be obtained by fitting the model to measured depth dose curves in order to describe individual characteristics of the beamline – the most important being the initial energy spread. We find that the free parameters of the depth dose model can be predicted for any intermediate energy from a couple of measured curves.

**Keywords:** proton treatment planning, Bethe-Bloch, Bragg peak, depth dose model, Landau tail, secondary reaction protons

## 1. Introduction

Radiotherapy with protons is becoming a modality with increasing importance, which triggers a lot of work in the field of algorithms for treatment planning. One of the challenges in treatment planning in general is to find a reasonable compromise between the speed and the accuracy of an algorithm. The fastest dose calculation algorithms are based on look-up tables for depth dose and lateral distributions of spread-out Bragg peaks or single pristine Bragg peaks (Hong et al 1996, Petti 1992, Deasy 1998, Schaffner et al 1999, Russel et al 2000, Szymanowski et al 2002, Ciangaru et al 2005). The look-up tables often consist of measured data directly or an analytical model fitted to the measurements in water. A depth scaling procedure is always applied in order to convert from water to another medium. Some authors use higher order corrections or different calculation approaches for the lateral distribution. Another approach is an iterative numerical calculation of the dose deposited by a proton beam along its path through the medium (Sandison et al 1997, Hollmark et al 2004). Monte Carlo methods can lead to the highest accuracy results – especially in highly heterogeneous media – but can still not be performed with sufficient speed for routine treatment planning (e.g. Petti 1996, Jiang et al 2004, Tourovsky et al 2005).

Common to most of the cited papers is that an adaptation procedure of the dose model to different properties of the beamline is not clearly elaborated. Some of the models may not be easy to adapt to another beamline or an intermediate range/energy at all. An exception to this is the model published by Bortfeld (1997). However, this model does not describe the buildup observed in higher energy protons and the transport of secondary protons. Furthermore, we present a theoretical analysis of the origin of the buildup effect, which is measured in higher energy proton beams. Due to the insufficient transport of secondary protons, this buildup is not modeled correctly by the Monte Carlo code PTRAN, which is used by other authors for a comparison with their models (e.g. Carlsson et al 1997, Sandison et al 2000).

The present model, which is most widely implemented in the commercial treatment planning package Eclipse[1], is built upon proton beamlets in water (Eclipse$^{TM}$ 2008 User Manual accessible with permission of Varian Inc.). They are a 3D dose distribution delivered by the quasi-monoenergetic beam impinging on a water surface with no lateral extension and angular divergence (by quasi-monoenergetic beam we refer to the spectrum as produced by the accelerator and beamline without any intended range modulation). The beamlet can be separated into a depth dose component and a lateral distribution. The depth dose obtained from a quasi-monoenergetic beam is often called the pristine Bragg peak. The present work focuses on the theoretical depth dose model that contains some fitting parameters - as defined below - which are used to fit our model to measured pristine Bragg peaks to account for beamline specific characteristics of our model. We will show how well the model can be adapted to different measured Bragg peaks – including buildup effects. The measurements have been obtained from different accelerators and will be described in more detail in the appropriate sections. The lateral component of the beamlet only includes scattering in the patient. It does not require any beamline specific configuration. The beamline specific component of the lateral penumbra is modeled through the lateral distribution of the in-air fluence (Schaffner 2008). In the development of the model, we have used GEANT4 (GEANT4 documents 2005) to analyze lateral scattering distributions, Landau tails, the production of heavy recoil particles and to obtain some numerical parameters. Many comparisons between Monte Carlo results and an exact integration of BBE, and nuclear cross-sections, respectively, have been previously described (Ulmer 2007).

**2. Material and methods**

In order to reduce the length of the study, we have added a Laboratory Report referred to as LR (and an L preceding references to equations, figures and tables). This LR is available online of this journal and provides necessary information on other publications (Ulmer 2007, Ulmer and Matsinos 2010) used in this study.

*2.1 Range – energy relations and integration of BBE*

An essential aspect in therapy planning is the relationship between initial proton energy $E_0$ and the range of the continuous-slowing-down-approximation ($R_{CSDA}$). The subsequent equations for $R_{CSDA}$, $E(z)$ and $dE(z)/dz$ have been previously derived (Ulmer 2007), and by integration of BBE (Bethe 1953, Bethe et al 1953, Bloch 1933, Boon 1998, ICRU49 1993, Segrè 1964) we have obtained the useful equation (section LR.1 of LR):

$$R_{CSDA} = \frac{1}{\rho} \cdot \frac{A_N}{Z} \sum_{n=1}^{N} \alpha_n E_I^{p_n} E_0^n \quad (\lim N \Rightarrow \infty) \quad (1)$$

---
[1] Varian Medical Systems Inc.

$E_I$ refers to an averaged ionization energy, $Z/A_N$ to the nuclear charge/mass number of the absorbing medium, and $\rho$ to its density (g/cm$^3$). For water we used $Z = 10$, $A_N = 18$, $\rho = 1$ g/cm$^3$ (Bragg rule) and $E_I = 75.1$ eV. Then Eq. (1) becomes:

$$R_{CSDA} = \sum_{n=1}^{N} a_n E_0^n \quad (\lim N \Rightarrow \infty) \quad (2)$$

N = 4 yields very accurate results for energies below 300 MeV. Parameters of Eq. (1 – 2) are stated in Tables L1 and L2 (LR.1).

Eq. (2), valid for water, can also be transformed to a sum of exponential functions (see Eq. (L2)). By that, it can be inverted (i.e. calculation of the (input) energy from the residual range: $E = E(R_{csda} - z)$). A restriction to N = 5 leads to a very high accuracy for proton energies below 300 MeV.

$$\left. \begin{aligned} E_0 &= R_{csda} \cdot \sum_{k=1}^{5} A_k \cdot e^{-R_{csda}/\beta_k} \\ E(z) &= (R_{csda} - z) \cdot \sum_{k=1}^{5} A_k \cdot e^{-(R_{csda}-z)/\beta_k} \end{aligned} \right\} \quad (3)$$

Parameters of Eq. (3) are stated in Table L3; formula (L4) and Table L3 also provide the inversion formula of Eq. (1) as a generalization of Eq. (3) and valid for arbitrary media.

Since $E(z)$ is analytically known, the stopping power $S(z) = dE(z)/dz$ can be derived from $E(z)$:

$$S(z) = \frac{dE(z)}{dz} = \sum_{k=1}^{5} A_k \cdot [\beta_k^{-1} \cdot (R_{CSDA} - z) - 1] \cdot e^{-(R_{CSDA}-z)/\beta_k} \quad (4)$$

Eq. (4) consists of a sum of 5 exponential functions. In order to speed up this expression with respect to dose calculations (50 MeV: factor 2.4 up to 250 MeV: factor 4.7; in connection with Gaussian convolutions the accelerated algorithm is a factor 2 - 3 faster, again), we reduce the 5 exponential functions of $S(z)$ to a single one of type (4) and introduce 4 more convenient (much less time-consuming) functions according to the following criteria of Eq. (4) and leading to $S_{approx}$ with the functions $\varphi_1 \ldots \varphi_5$ in Eq. (5):

1. Optimization of the exponential behavior and coefficient weight, slightly depending on the initial energy $E_0$ by an envelope exponential function $\exp(-Q_p \cdot (R_{CSDA} - z))$, to provide the main contribution of the exponentially increasing part of Bragg curves ($\varphi_3$).

2. A Gaussian term ($\varphi_1$) containing a half-width $\tau_0 \approx 10^{-5}$ cm aims the reflecting behavior of the Bethe-Bloch function in the environment of the CSDA-range, which would otherwise be singular and could not be integrated. Thus, we are using $\exp(-(R_{CSDA} - z)^2 / \tau_0^2)$ instead of the δ-function (if $\lim \tau_0 \to 0$); in the subsequent Eq. (6), the undefined square of a δ-function would appear instead of products with Gaussian terms resulting from $\varphi_1$. The problem of the singularity does not exist anymore in the presence of range straggling as represented by the parameter τ, which will be defined in a later section. Later on, $\tau_0$ will therefore be neglected ($\tau_0 \ll \tau_{straggle}$).

3. A power expansion of Eq. (4) with respect to the initial plateau and slowly increasing domain of $S(z)$ up to the order $z^2/R_{CSCA}^2$ provides the functions $\varphi_2$, $\varphi_4$ and $\varphi_5$.

With the help of these 5 functions we are able to develop an accelerated algorithm:

$$\left. \begin{array}{l} 1.\ \varphi_1 = C_1(E_0) \cdot \exp(-(R_{CSDA} - z)^2 / \tau_0^2) \cdot \theta(R_{CSDA} - z) \\ 2.\ \varphi_2 = 2 \cdot C_2(E_0) \cdot \theta(R_{CSDA} - z) \\ 3.\ \varphi_3 = 2 \cdot C_3(E_0) \cdot \exp(-Q_p(E_0) \cdot (R_{CSDA} - z)) \cdot \theta(R_{CSDA} - z) \\ 4.\ \varphi_4 = 2 \cdot C_4(E_0) \cdot (z/R_{CSDA})^2 \cdot \theta(R_{CSDA} - z) \\ 5.\ \varphi_5 = 2 \cdot C_5(E_0) \cdot (1 - z/R_{CSDA}) \cdot \theta(R_{CSDA} - z) \end{array} \right\} \quad (5)$$

Explanations: $\theta(R_{CSDA} - z)$ is a unit step function, i.e. $\theta(R_{CSDA} - z) = 1$ (if $z \leq R_{CSDA}$) and 0 (otherwise). The purpose of the unit step function is that the energy $E(z)$ is zero for $z > R_{CSDA}$. $Q_p = \pi \cdot P_E/z_{max}$ appears in the function $\varphi_3$; $z_{max}$ will be explained in section 2.3. The parameter $P_E$ and the coefficients $C_1$, $C_2$, $C_3$, $C_4$, and $C_5$ depend linearly on $E_0$ and are determined by the variation procedure:

$$\left. \begin{array}{l} \sum_{E_0=1}^{300} \int_0^{R_{CSDA}} |S(z) - S_{approx}(z)|^2 dz = Minimum \\ \\ S(z) \approx S_{approx} = \sum_{k=1}^{5} \varphi_k(z, E_0) \end{array} \right\} \quad (6)$$

It should be mentioned that the determination of the stopping power function according to Eq. (5) and Eq. (6) agrees with an internally considered expansion of the solution functions of BBE with Feynman propagators, but the energy-dependence of the parameters had to be defined by GEANT4. In particular, the free particle propagator is also a Gaussian kernel, and $\varphi_1$ immediately results from this kernel; the remaining contributions 2 – 5 of Eq. (5) are generated via iterated integral operators. Since an analytical integration of BBE is superior to a perturbation expansion based on propagators, we prefer to present here the solution method without propagators. The result of the adaptation of S(z) according to Eq.(6) is stated in Table 1. The mean standard deviation amounts to 0.7 %. It turned out that the contribution of the coefficient $C_5$ is rather negligible, since it amounts to 0.007, and so this contribution may be omitted ($C_5 = 0$). Therefore the backbone of the accelerated algorithm can be restricted to the four coefficients $C_p$ (p = 1,…, 4) and a parameter $P_E$ related to $\varphi_3$. The linearly energy-depending $C_1,…, C_4$ and $P_E$ are given by:

$$C_p = \alpha_{0,p} + \alpha_{1,p} \cdot E_0 \ (p=1,..,4) \ and \ P_E = \alpha_{0,5} + \alpha_{1,5} \cdot E_0 \qquad (7)$$

Table 1: Parameters to calculate energy dependence of the coefficients $C_p$ and $P_E$ according to Eq. (7)

| $C_p$ | $C_1$ | $C_2$ | $C_3$ | $C_4$ | $P_E$ |
|---|---|---|---|---|---|
| α0,p | 2.277463 | 0.2431 | 1.0295 | 0.4053 | 6.26751 |
| α1,p | - 0.0018473 | 0.0007 | - 0.00103 | - 0.0007 | 0.00103 |

A determination of the parameters presented in Table 1 with GEANT4 (least-square fit) leads to the following maximal deviations: $C_1$: + 0.07 %; $C_2$ = + 0.08 %; $C_3$: +0.04 %; $C_4$: -0.09 %; $P_E$: -0.08 % at different values of $E_0$. The mean standard deviations are of the order 0.04 % – 0.06 %. A generalization of Eq. (7) from water to arbitrary media is given in Eq. (L5) and Table L3 in LR.1.

### 2.2 The fluence decrease of primary protons due to nuclear interactions

According to an investigation presented previously (Ulmer 2007) the decrease of the fluence of primary protons $\Phi_{pp}$ due to nuclear interactions of protons in water can be described by:

$$\left. \begin{array}{c} \Phi_{PP}(z) = \frac{1}{2}\left(1 - uq \cdot \frac{z}{R_{csda}}\right) \cdot \left[1 + \mathrm{erf}\left(\frac{R_{csda} - z}{\tau}\right)\right] \cdot \Phi_0 \\ uq = [(E_0 - E_{Th})/M_p c^2]^{1.032} \end{array} \right\} \qquad (8)$$

$\Phi_0$ is the initial fluence of protons (dimensionless), $E_0$ the initial proton energy, $E_{Th}$ = 7 MeV, and $M_p c^2$ = 938.276 MeV is the proton rest energy. Eq. (8) results from an integration of the total nuclear proton – oxygen cross-section (Figures L2 – L3, data of Chadwick et al 1996, and calculations of Ulmer and Matsinos 2010). $E_{Th}$ = 7 MeV is the threshold energy necessary to surmount the Coulomb repulsion of the oxygen nucleus. $\Phi_0$ represents an arbitrary initial fluence of a proton beamlet at surface.

It might be surprising that in formula (8) the error function *erf* and the rms-value τ of a Gaussian appear. In principle, the behavior of $\Phi_{pp}$, valid within the CSDA-framework, should be a straight line as long as E = $E_{Th}$ is not yet reached. For E < $E_{Th}$ to E = 0, $\Phi_{pp}$ should be constant; at E = 0 (z = $R_{CSDA}$) a jump to $\Phi_{pp}$ = 0 is expected. However, due to energy/range straggling, the proton beam can never remain monoenergetic in the sense of CSDA. Since τ refers to the half-width of a Gaussian convolution, we introduce 'roundness' in the shape. The range of 7 MeV protons is less than 1 mm; therefore, we cannot verify whether $\Phi_{pp}$ is constant in the fluence profile of primary protons. The half-width parameter $\tau$ will be defined subsequently.

At first, we adopt the commonly applied assumption that energy/range straggling can be described by a Gaussian[2] type of fluctuation. This means, that we can add the width of the corresponding distributions quadratically, i.e., $\tau$ is given by:

$$\tau = \sqrt{\tau_{straggle}^2 + \tau_{in}^2} \qquad (9)$$

The parameter $\tau_{in}$ represents the distribution of the incident beam and $\tau_{straggle}$ the variation of the range due to straggling along the beam path. It reaches its maximum at $R_{CSDA}$:

$$\tau_{straggle}(R_{CSDA}) = \sqrt{2} \cdot 0.012703276 \cdot \begin{cases} R_{csda}^{0.9352}, & \text{if } R_{csda} \geq 1\,\text{cm} \\ R_{csda}^{1.763}, & \text{if } R_{csda} < 1\,\text{cm} \end{cases} \qquad (10)$$

For most applications, it is sufficient to use a constant value for $\tau_{straggle}$ along the whole beam path. However, in some cases, we want to use the depth dependence of $\tau_{straggle}$ (see section 2.6 about Landau tail corrections). It should be noted (Ulmer 2007), that $\tau_{straggle}$ increases exponentially with the stopping power function $S(z)$ according to BBE:

$$\left.\begin{aligned}\tau_{straggle}(z) &= \tau_{straggle}(R_{CSDA}) \cdot \frac{e^{Q_z \cdot z} - 1}{e^{Q_z \cdot R_{CSDA}} - 1} \\ Q_z &= 2.887\,cm^{-1}\end{aligned}\right\} \qquad (11)$$

*2.3  The stopping power of primary protons*

The overall calculation of the depth dose deposited by primary protons follows from a combination of the fluence reduction (section 2.2 and the results of the integration BBE (section 2.1)). The stopping power of primary protons thus becomes:

$$S_{pp}(z,E_0) = \Phi_0 \cdot \left(1 - \frac{uq \cdot z}{R_{csda}}\right) \cdot \left[I_1(z,E_0) + I_2(z,E_0) + I_3(z,E_0) + I_4(z,E_0) + I_{Lan1}(z,E_0) + I_{Lan2}(z,E_0)\right] \qquad (12)$$

$\Phi_0$, which has to be dimensionless and can be put 1, and the following factor $(1 - uq\cdot z/R_{csda})$ represent the number of remaining protons at the given depth. The 'I-terms $I_1 - I_4$' result from functions $\varphi_1 - \varphi_4$ of Eq. (5) subjected to a Gaussian convolution with $\tau$ according to Eq. (9) to account for energy/range straggling. The terms $I_{Lan1}$ and $I_{Lan2}$ result from the fact that energy fluctuations are not symmetrical, as required by the Gaussian kernel. For the sake of clearness we shall treat the contributions of Landau tails $I_{Lan1}$ and $I_{Lan2}$ separately (section 2.6). The remaining terms $I_1(z, E_0) - I_4(z, E_0)$ are determined by:

$$I_1 = \frac{1}{N_{abs}} \cdot \left( C_1 \cdot \frac{\tau_{straggle} \cdot R_{csda}}{\tau} - C_4 \cdot \frac{\tau \cdot (R_{csda} + z)}{\sqrt{\pi} R_{csda}^2} \right) \cdot e^{\left(-\frac{(R_{csda} - z)^2}{\tau^2}\right)} \quad (13)$$

$$I_2 = \frac{1}{N_{abs}} \cdot \left( C_2 + C_4 \cdot \frac{\tau^2}{2\sqrt{\pi} R_{csda}^2} \right) \cdot \left( 1 + erf\left( \frac{R_{csda} - z}{\tau} \right) \right) \quad (14)$$

$$I_3 = \left\{ \frac{1}{N_{abs}} \cdot C_3 \cdot \exp\left( \frac{(P_E \cdot \pi \cdot z_{max} \cdot \tau_{in})^2}{4} - \frac{P_E \cdot \pi \cdot (R_{csda} - z)}{z_{max}} \right) \cdot \left[ 1 + erf\left( \frac{R_{csda} - z}{\tau} - \frac{P_E \cdot \pi \cdot \tau}{2 \cdot z_{max}} \right) \right] \right\} \quad (15)$$

$$I_4 = \frac{1}{N_{abs}} \cdot C_4 \cdot \frac{z^2}{R_{csda}^2} \cdot \left[ 1 + erf\left( \frac{R_{csda} - z}{\tau} \right) \right] \quad (16)$$

---

[2] The usual form of the Gaussian involves the expression $\exp(-x^2/(2\sigma^2))$ For computational reasons, we substitute $2\sigma^2$ by $\tau^2$ and indicate this by using $\tau$, where referring to the Gaussian width.

The parameters $C_1...C_4$ and $P_E$ can be calculated from Eq. (7) and Table 1. The normalization factor $N_{abs}$ depending on $R_{csda}$ will be stated in section LR.5. For fast relative calculations in therapy planning (normalization of the Bragg peak: 1 or 100 %) $N_{abs}$ can be put 1; only for absolute calculations (Gy/MU, see section 3.1) the real value of $N_{abs}$ is required (Eclipse$^{TM}$ 2008). It should be noted that $\tau$ refers to Eq. (9), but a restriction to $\tau_{in} = 0$ transforms equations (13 - 16) to the monoenergetic case. Parameter $z_{max}$, which appears in Eq. (15), is given by $z_{max} = R_{csda} + \tau_{Range}$; $\tau_{Range}$ is calculated by the formula:

$$\tau_{Range} = R_{CSDA} \cdot [2.117908559 \cdot 10^{-5} \cdot E_0 + 0.919238854 \cdot 10^{-7} \cdot E_0^2] \quad (17)$$

A further use of $\tau_{Range}$ can be verified in LR.4.

*2.4 The stopping power of secondary and recoil protons*

Before discussing the details of the modeling of secondary protons, we need to clarify the nomenclature. We consider all protons to be secondary protons, if they did undergo a nuclear interaction with the nucleus. The calculations in section 2.2 (decrease of primary proton fluence) is based on the total nuclear cross-section of the interaction proton with the oxygen nucleus (Figure L2 and similar figures in Medin et al 1997, Paganetti 2002, Boon 1998). This cross-section leads to a steady decrease of primary protons with a slope, which is in exact agreement with results of Boon (1998) and the Los Alamos Scientific Library. Figure L3 shows for example a 20% decrease of primary protons for an initial energy of 200 MeV protons. Based on GEANT4 Paganetti (2002) states similar numbers, but he and other authors (Fippel et al 2004, Medin et al 1997) still call those protons 'primary' protons, which we shall describe by the subsequent case 1. In our work, we classify all protons emerging from nuclear interactions as secondary protons. Looking at secondary protons in detail, we can distinguish two cases:

1. Secondary protons by potential/resonance scattering (*sp,n*: nonreaction protons). Potential scattering (elastic) results from the strong interaction potential in the environment of the nucleus (note that $R_{Strong} \approx 1.2 \cdot A_N^{1/3} \cdot 10^{-13}$ cm determines the distance with a balance between strong interaction force and Coulomb force; $A_N$(oxygen) = 16). For $R > R_{Strong}$ only the Coulomb part is present. Resonance scatter (inelastic) is a typical quantum mechanical effect; it results from the proton/nucleus interaction inducing transitions between different states of the nucleus (e.g. vibrations leading to intermediate deformations, rotation bands, excited states by changing the spin multiplicity, $E_{res}$ (oxygen): 20.12 MeV). These elastic and inelastic processes are described by the Breit-Wigner-Flügge formula (see e.g. Segrè 1964).

2. Nuclear reactions (*sp,r*: reaction protons, inelastic), which produce heavy recoils (see e.g. the listing (27)). These protons are sometimes referred to as reaction protons – or 'secondary protons' by the authors cited above. For protons in the therapeutic energy domain the contribution of reaction protons to the total dose amounts to 1 % (160 MeV) - 4 % (270 MeV) and increases with energy.

Within the total nuclear cross-section case 1 plays the dominant role, if E < 100 MeV. If E > 100 MeV the importance of case 2 surmounts case 1 and increases with the energy. In particular, the contribution of resonance scatter is drastically decreasing (Ulmer and Matsinos 2010). In both cases ca. 1 – 7 MeV have to be added to the proton energy (this depends on the deflection angle); they are transferred to whole nucleus to satisfy energy/momentum conservation in the center-of-mass system. This implies that for a neutron release the proton energy has to be 21 – 27 MeV and not simply 20 MeV (an exception is stated in the discussion of listing (27)). Due to potential barrier the energy of the colliding proton has, at least, to be 30 MeV in order to release a secondary proton.

The fluence of secondary protons $\Phi_{sp}$ ($\Phi_{sp,n}$ and $\Phi_{sp,r}$) and recoil protons/neutrons $\Phi_{rp}$ together is only approximately equal to the fluence loss of primary protons due to nuclear interactions (term (1-$uq \cdot z/R_{csda}$) in Eq. (8)). The relative weight between the two types was obtained from Monte Carlo simulations. We therefore obtain for the fluence of secondary nonreaction (*sp,n*) and recoil protons the following equations:

$$\Phi_{sp,n} = \upsilon(1 - 2 \cdot C_{heavy}) \cdot \left(\Phi_0 \frac{uq \cdot z}{R_{csda}}\right)$$

$$\Phi_{rp} = 0.042 \cdot \left(\Phi_0 \frac{uq \cdot z}{R_{csda}}\right) \qquad (18)$$

$C_{heavy}$ is determined according to Eq. (26). At this place, we already point out that the factor $\upsilon$ in Eq. (18) referring to the fluence of secondary protons has not simply to be 0.958, since it may depend on the beamline (see also section 3).

The transport of secondary and recoil protons according to Eq. (18) basically shows the same physical behavior as primary protons. Therefore, we can describe them with the same terms as primary protons (section 2.3):

$$S_{sp,n}(z_s, E_0) = \Phi_0 \cdot \left(\upsilon(1 - 2 \cdot C_{heavy}) \cdot \frac{uq \cdot z}{R_{csda}}\right) \cdot$$
$$[I_1(z_s, E_0) + I_2(z_s, E_0) + I_3(z_s, E_0) + I_4(z_s, E_0) + I_{Lan1}(z_s, E_0) + I_{Lan2}(z_s, E_0)] \qquad (19)$$

$$S_{rp}(z_s, E_0) = \Phi_0 \cdot \left(0.042 \cdot \frac{uq \cdot z}{R_{csda}}\right) \cdot$$
$$[I_1(z_s, E_0) + I_2(z_s, E_0) + I_3(z_s, E_0) + I_4(z_s, E_0) + I_{Lan1}(z_s, E_0) + I_{Lan2}(z_s, E_0)] \qquad (20)$$

However, due to energy loss of primary protons, the Bragg peak of secondary, nonreaction protons (*sp,n*) is shifted to a lower z-value. The depth coordinate z, therefore, has to be replaced by a shifted coordinate, i.e.

$$z \Rightarrow z_s = z + z_{shift} \qquad (21)$$

In addition, the convolution parameter $\tau$ also contains a term resulting from the nuclear cross-section:

$$\tau = \sqrt{\tau_{straggle}^2 + \tau_{in}^2 + 0.5 \cdot \tau_{heavy}^2} \qquad (22)$$

Practically, this means that secondary protons ('*sp, n*') alone also exhibit a Bragg peak, but one which is much broader than the Bragg peak of primary protons (see also Figure 2). Using previous results (Ulmer 2007) $z_{shift}$ and $\tau_{heavy}$ are given by:

$$z_{shift} = \begin{cases} 0 & (if\ E_0 < E_{res}) \\ \sum_{n=1}^{4} a_n E_{res}^n \cdot [1 - \exp\left(-\frac{(E_0 - E_{res})^2}{5.2^2 \cdot E_{res}^2}\right)] & (if\ E_0 \geq E_{res}) \end{cases} \qquad (23)$$

$$\tau_{heavy} = \begin{cases} 0.00000, & \text{if } E_0 < E_{Th} \\ 0.55411 \cdot \dfrac{E_0 - E_{Th}}{E_{res} - E_{Th}}, & E_{Th} \leq E_0 \leq E_{res} \\ 0.554111 - 0.000585437 \cdot (E_0 - E_{res}), & \text{if } E_{res} < E_0 \end{cases} \quad (24)$$

*2.5 The contribution of heavy recoils and their connection to secondary/recoil protons*

Finally, we have to complete the total stopping power function S(z) for the reaction proton (*sp,r*) and heavy recoil contribution. With regard to heavy recoils an analysis of Monte Carlo calculations (GEANT4) for protons travelling in water for energies from 5 MeV up to 270 MeV showed that it is possible to fit the data with the help of the following equations:

$$S_{heavy}(z, E_0) = \Phi_0 \, C_{heavy} \cdot e^{-\frac{z}{z_{max}}} \cdot \left[ 1 + erf\left( \sqrt{2} \cdot \frac{R_{csda} - z + \frac{E_0}{250} - 1}{\tau_{heavy}} \right) \right] \quad (25)$$

$$C_{heavy} = \begin{cases} 0 & (\text{if } E_0 < E_{Th}) \\ 0.00000042643926 \cdot (E_0 - E_{Th})^2 & (\text{if } E_0 \geq E_{Th)}) \end{cases} \quad (26)$$

With respect to the threshold energy $E_{Th}$ = 7 MeV we have been able to verify a very small 'jump' depending on the available proton energy, i.e. there is no continuous transition from $E_0 < E_{Th}$ to $E_0 = E_{Th}$. The most probable heavy recoil elements resulting from the nuclear reactions by therapeutic protons are the cases 1 – 5; cases 6 – 7 result from cases 1 - 2:

$$\begin{aligned}
&1.\, p + O_8^{16} \Rightarrow n(neutron) + F_9^{16} \, (\beta^+ - decay, T_{1/2} = 22 \text{ sec} + \gamma) \\
&2.\, p + O_8^{16} \Rightarrow p + n + O_8^{15} \, (+\beta^+ - decay, T_{1/2} = 124 \text{ sec} + \gamma) \\
&3.\, p + O_8^{16} \Rightarrow p + p + N_7^{15} \, (+\beta^+ - decay, T_{1/2} = 10 \text{ min} + \gamma) \\
&4.\, p + O_8^{16} \Rightarrow \alpha + N_7^{13} \, (+\beta^+ - decay, T_{1/2} = 124 \text{ sec} + \gamma) \\
&5.\, p + O_8^{16} \Rightarrow d + O_8^{15} \, (+\beta^+ - decay, T_{1/2} = 124 \text{ sec} + \gamma) \\
&6.\, n + O_8^{16} \Rightarrow p + N_7^{16} \, (+\beta^- - decay, T_{1/2} = 120 \text{ sec} + \gamma) \\
&7.\, n + O_8^{16} \Rightarrow p + n + N_7^{15} \, (see \text{ case } 3)
\end{aligned} \quad (27)$$

All types of β⁺-decay emit one γ-quant; its energy is of the order 0.6 MeV – 1 MeV. The β⁺-decay of $F_9^{16}$ has a half-life of about 20 seconds, and further γ-quanta are produced by collisions of positrons with environmental electrons. The remaining heavy recoil fragments have partially half-times up to 10 minutes ($N^{15}$). Since Figure L2 refers to the total nuclear cross-section in dependence of the actual (residual) proton energy, we have to add some qualitative aspects on the 5 different types with regard to the required proton energy: If E < 50 MeV the type (1) is the most probable case with rapid decreasing tendency between 50 MeV < E < 60 MeV to become zero for E > 60 MeV. Type (2) also pushes out a neutron, but the incoming proton is not absorbed; the required energy amounts, at least, to 50 MeV. Type (3) is similar, but requires, at least, ca. 60 MeV with increasing probability. The release of α-particles resulting from clusters in the nucleus requires E ≈ 100 MeV and the probability is increasing up to E ≈ 190 MeV; thereafter it is decreasing rapidly, since higher energy protons destroy these clusters by pushing out deuterons according to type 5. Thus case 5 is energetically possible for E > 60 MeV, but the significance is only increasing for E > 200 MeV. The nuclear reactions 6 and 7 result from the release of neutrons; they may undergo further interactions with oxygen. According to Ulmer and Matsinos 2010) the cases 1 and 6 are noteworthy. Thus the neutron released by the incoming proton has not absolutely to be the result of a real collision (threshold energy, at least, 20 MeV). For energies E > 7 MeV and E < 20 MeV a resonance effect via exchange interaction between proton and nucleus via a $\pi^-$ meson (Pauli principle) is also possible. By that, the incoming proton leaves the oxygen nucleus as a neutron. Case 6 represents the reversal process, i.e., the incoming (secondary) neutron is converted to a proton via $\pi^+$ exchange. *The calculation procedure for $S_{sp,r}$ is presented in LR, section LR.2* (see also Ulmer and Matsinos 2010).

We should also point out that nuclear reactions like relations (27) are only a part of the total decrease of the primary proton fluence according to Eq. (8). If E < 100 MeV, the main part of the decrease of the primary proton fluence results from nuclear scatter of protons by the oxygen nucleus (deflection of primary protons without release of further nucleons) via intermediate deformations and oscillations of the nucleus. These oscillations are damped by emission of γ-quanta with very low energy (ca. 1 keV), which are most widely absorbed by the Auger effect. The most important source for recoil protons are elastic collisions of projectile protons with the proton of Hydrogen. Released neutrons of type 1 also lose most widely their energy by such collisions before they become thermal neutrons. These neutrons usually escape and undergo $\beta^-$ - decay to produce a proton, electron and a γ-quant (0.77 MeV), $T_{1/2}$ = 17 min. Neutrons of type 2 carry a much higher energy and can escape without any significant collisions. The cases 6 and 7 referring to the neutron interaction with the oxygen nucleus are the only noteworthy inelastic contributions. Fig. 1 shows the dose contribution of reaction protons for some therapeutic proton energies. The contributions of deuterons and α-particles are also accounted for in this figure. The tails beyond the ranges $R_{CSDA}$ mainly result from the reaction types 6 and 7 of the listing (27).

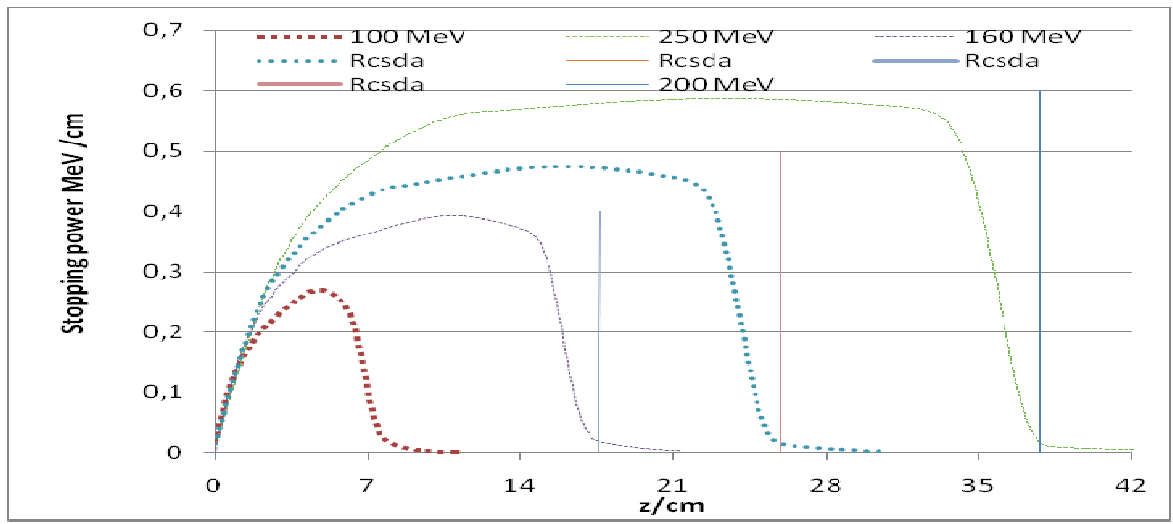

Fig. 1. Stopping power of secondary/tertiary protons for the initial proton energies 100 MeV, 160 MeV, 200 MeV and 250 MeV. The $R_{CSDA}$ ranges are indicated by perpendicular straight lines.

*2.6   Landau tail corrections and their influence to the buildup of higher energy proton beams*

As already indicated, we have not yet determined the contributions $I_{Lan1}$ and $I_{Lan2}$ resulting from the modification of the energy transfer and stopping power of protons by corrections resulting from Landau tails. A theoretical analysis (Ulmer 2007) yields that symmetrical – i.e. Gaussian - fluctuations (and related convolution kernel) of the energy transfer according to the continuous-slowing-down-approximation (CSDA) are only rigorously valid, if the local proton energy and the energy transfer by collisions are nonrelativistic.  The maximum energy transfer $E_{max}$ as a function of the local energy for protons in water is shown in Figure L6. $E_{max}$ has a non-linear term (relativistic correction), which becomes more important with increasing energy. In a similar way, the fluctuations of the energy transfer become less symmetrical; i.e. collisions occur much less frequent. This behavior can be observed more and more for proton energies greater than about 100 MeV. A consequence of this relativistic effect is that protons in the incident region (e.g. 250 MeV, $E_{max}$ = 617 keV) undergo less hits and, on the other side, the energy transfer per hit to environmental electrons is much higher than expected from a symmetrical energy transfer. Thus, the energy transfer of 250 MeV protons behaves in a manner such as if protons with E > 250 MeV or >> 250 MeV would transfer energy in a symmetrical way (in the initial plateau region, a Gaussian convolution does not change the Bragg curve obtained by CSDA, but the stopping power at surface is reduced (ICRU49)).   A buildup effect can be seen as long as the symmetrical fluctuation is not yet reached. This effect decreases along the proton track, and when the local energy approaches about 100 MeV, the fluctuations of the energy transfer tend to become symmetrical and the buildup effect is negligible.

A relativistic treatment (Dirac equation and Fermi-Dirac statistics - section LR.3) of statistical fluctuations of the energy transfer yields that the inclusion of Landau tails (this leads to a Vavilov distribution function) implies modified Gaussian convolutions, i.e. a Gaussian convolution with relativistic correction terms expressed by two-point Hermite polynomials. Thus, the Gaussian convolution kernel additionally contains a series expansion of Hermite polynomials; but for proton energies below about 300 MeV lower order corrections up to second order $H_2$ are sufficient. The results of these corrections are the following terms:

$$\left. \begin{array}{l} I_{Lan1} = \frac{1}{2} \cdot \frac{C_{Lan1}}{N_{abs}} \cdot \left\{ erf\left(\frac{z'}{\tau_{Lan1}}\right) + erf\left(\frac{(R_{CSDA} - z')}{\tau_{Lan1}}\right) \cdot z / R_{CSDA} \right\} \\ C_{Lan1} = -1.427 \cdot 10^{-6} \cdot R_{csda}^{3} + 1.439 \cdot 10^{-4} \cdot R_{csda}^{2} - 0.002435 \cdot R_{csda} + 0.2545 \\ \tau_{Lan1} = (R_{csda} \cdot (0.812087912 - 0.001648352 \cdot E_0) + 0.492 \cdot \tau_{in}) / 2 \\ z_{Ls} \ (in \ cm) = 0.602 \cdot E_0 / 270, \ z' = z + z_{Ls} \end{array} \right\} \quad (28)$$

Please note that z' and consequently $z_{Ls}$ in Eq. (29) agree with Eq. (28).

$$\left. \begin{array}{l} I_{Lan2} = -\frac{C_{Lan2}}{N_{abs} \cdot R_{Lan2}^{2}} \left\{ \left( R_{Lan2}^{2} - z^2 - \frac{\tau_{Lan2}^{2}}{2\sqrt{\pi}} \right) \cdot \left( erf\left(\frac{z'}{\tau_{Lan2}}\right) + erf\left(\frac{R_{Lan2} - z'}{\tau_{Lan2}}\right) \right) + \frac{(z + R_{Lan2}) \cdot \tau_{Lan2}}{\sqrt{\pi}} \cdot e^{-\frac{(R_{Lan2} - z')^2}{\tau_{Lan2}^{2}}} \right\} \\ C_{Lan2} = \begin{cases} 0.00013059 \cdot E_0, & if \ E_0 \leq 168 \, MeV \\ 0.022, & if \ E_0 > 168 \, MeV \end{cases} \\ R_{Lan2} = (3.19 + 0.00161 \cdot E_0) \cdot \left[ 1 - \exp\left(-\frac{E_0^{2}}{165.795268^2}\right) \right]; \ \tau_{Lan2} = 2.4 \cdot \sqrt{\tau_{straggle}(z)^2 + \tau_{in}^{2}} \end{array} \right\} \quad (29)$$

The contribution terms $I_{Lan1}$ and $I_{Lan2}$ have to be accounted for in the previous sections referring to $S_{pp}$, $S_{sp,n}$ and $S_{rp}$. Equations (28 – 29) have to be used in the modified form $z \rightarrow z_s = z + z_{shift}$ and $z' \rightarrow z_s' = z' + z_{shift}$ with regard to the terms $S_{sp}$ and $S_{rp}$. A neglect of $z_{shift}$ in $S_{rp}$ enables us to add $S_{rp}$ to $S_{pp}$. For proton energies larger than about 120 MeV, the correction term $I_{Lan2}$ is of increasing importance. If the energy is lower than 120 MeV, the contribution $I_{Lan1}$ still remains noteworthy due to impinging polychromatic protons. The beamline of protons also shows a tail in the energy spectrum due to the nozzle, range modulator wheel, and inclusion of reaction protons. It is certainly not sufficient to take account for all these influences by a half-width parameter $\tau_{in}$ in a Gaussian convolution. $\tau_{in}$ also plays a role in Landau tails. The philosophy of Eclipse™ (2008) is to reduce fitting parameters as much as possible:

1. With regard to $I_{Lan2}$ we only have assumed that $\tau_{Lan2}$ remains consistent, if in the polychromatic case $\tau_{in}$ has the same weight as $\tau_{straggle}$. Possible fittings of $C_{Lan2}$ and $R_{Lan2}$ turned out to be superfluous.

2. Since the influence of the beamline depends on specific properties of the accelerator installation, we have implemented the adaptation of the parameters $I_{Lan1}$ by a fitting procedure in addition to the fitting via $\tau_{in}$. According to Eq. (28) the weight of $C_{Lan1}$ only represents an initial value, which may slightly be modified by fitting procedures to adapt machine-specific properties with regard to tails (section 'Results'). The weight of $C_{Lan1}$ is expected to be affected by the reaction protons of the beamline; this flexibility is necessary, since with regard to measured impinging protons we cannot distinguish between primary and secondary protons produced along the beamline (nozzle, etc).

The theoretically evaluated correction terms to account for Landau tails has also been subjected to comparisons with results of GEANT4 with respect to monoenergetic protons, i.e. $\tau_{in} = 0$, in steps of 2 MeV, beginning with 20 MeV up to 250 MeV. Since GEANT4 offers the possibility to include either Vavilov distribution functions or the restriction to a pure Gaussian fluctuation and to switch on or off the Hadronic generator, a comparison yielded that the difference between the calculation model according to equations (28 – 29) and Monte Carlo never exceeded 2.2 %, and the mean standard deviation amounted to 1.3 %. Referring to polychromatic proton spectra induced by the beamline, the sole check is given by a comparison with experimental data. It should be noted that the contribution to the buildup induced by Landau tails could also be verified by Monte Carlo. If Landau tails were taken into account, the buildup was increased and reduced, when statistical fluctuations were restricted to a Gaussian kernel. Figures L7 and L8 in LR.3 present some results of theoretical and GEANT4 calculations for monoenergetic primary protons (150 MeV, 200 MeV and 250 MeV). These examples clearly show the increasing role of Landau tails with proton energy. A partially different interpretation of buildup effects by secondary protons has previously been given (Carlsson et al 1977)). Opposite to our results, they interpreted this effect as the *only* result of secondary proton generation (*sp,r*) along the proton track (see e.g. Fippel et al 2004, Medin el al 1997 and Paganetti 2002, where this opinion has been adopted). Fig. 2 presents the total stopping power S and its partial contributions $S_{pp} + S_{rp}$, $S_{sp,n}$, $S_{sp,r}$ and $S_{heavy}$ for monoenergetic protons (250 MeV). Fig. 3 refers to Fig. 2 in the buildup region. The role of the reaction protons and Landau tail with regard to the buildup can be verified in Fig. 4. Thus for *monoenergetic* proton beams the reaction protons (*sp,r*) contribute ca. 65 % to the total buildup. This situation may be rather different for polychromatic proton beams, which we mainly characterize be the parameter $\tau_{in}$. In our model $\tau_{in}$ enters both, the broadening of the Gaussian rms-value as well as the Landau tail in the buildup region, expressed by the calculation of $\tau_{Lan1}$ and $\tau_{Lan2}$ in Eq. (29).

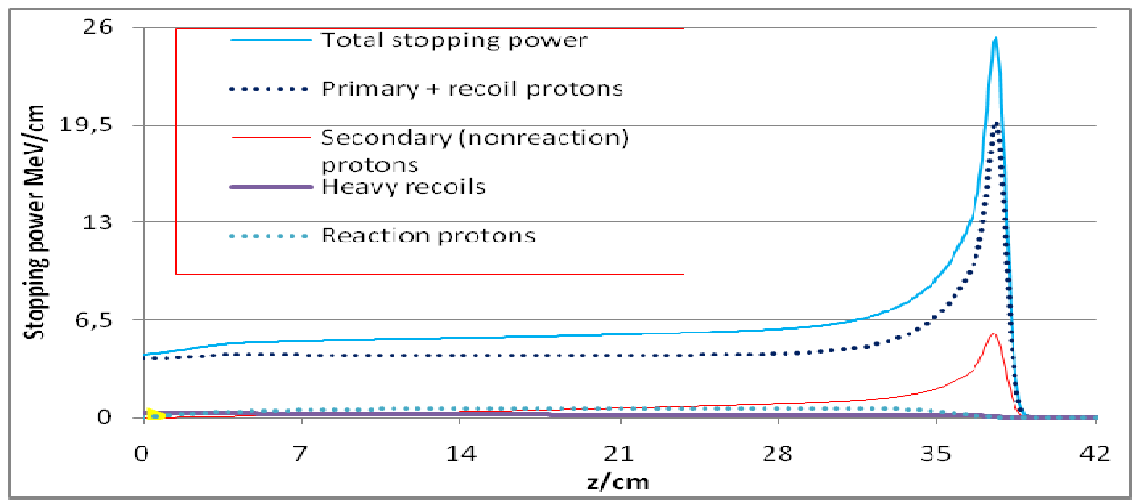

Fig. 2. Total stopping power S and related partial contributions of a monoenergetic 250 MeV proton beam.

If the impinging proton beam is additionally modulated by collimator scatter, the profiles of Bragg curves may be drastically changed by the collimator-surface distance and field-size (Hong et al 1996, Ulmer and Schaffner 2006). At every case, the parameter $\tau_{in}$ is used to model both, the width of the Bragg peak region and the role of all protons in the buildup region.

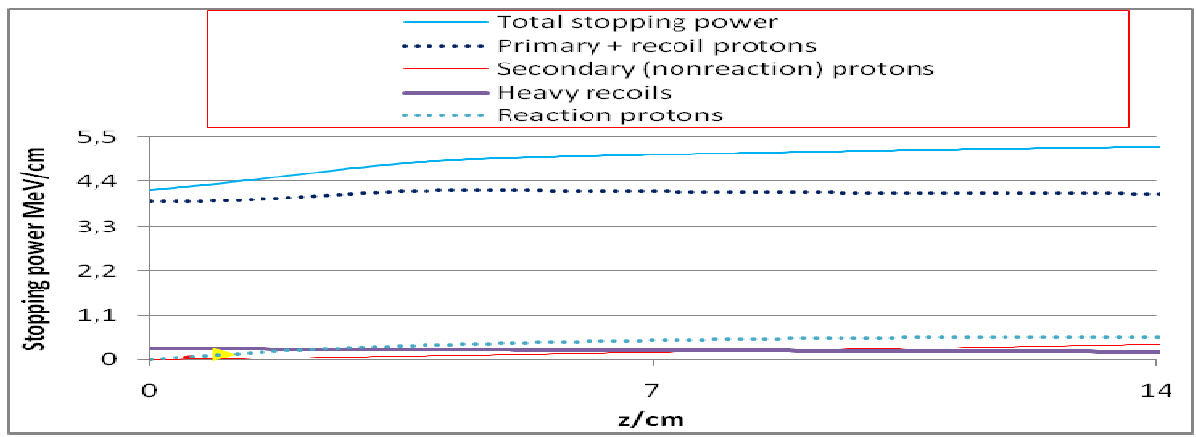

Fig. 3. Buildup region of Fig. 2.

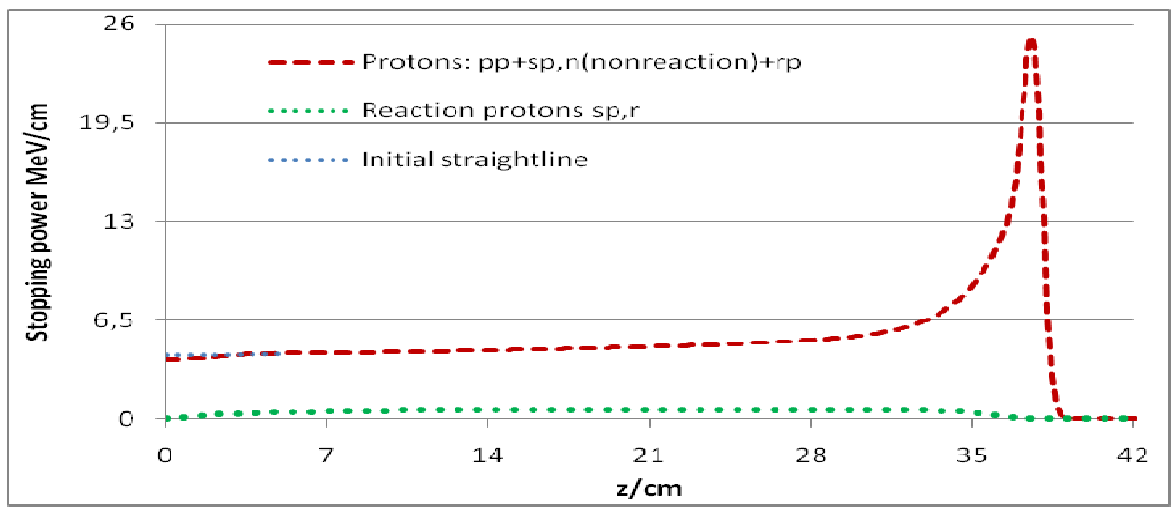

Fig. 4. Total contribution of all protons of Fig. 2,, excepted reaction protons. The buildup is decreased; this is indicated by the initial straight line.

We should note that collimators (blocks) may not only have an impact to transverse profiles, but also to the initial plateau of Bragg curves, if collimator-scattered protons cross behind the impinging plane. However, these effects depend on a lot of specific factors, e.g. collimator distance, SAD, field-size and collimator properties, e.g. used alloy and its length (see Ulmer and Schaffner 2006, Matsinos 2008).

*2.7 Calculation of 3D beamlets and 3D dose*

This calculation procedure has to be based on the knowledge of stopping power and lateral scattering functions. The structure of the scatter kernel $K_{lat}$ consists of two different kernels to treat primary/recoil/nonreaction protons and, on the other side, reaction protons and heavy recoils. The determination of these two kernels is developed in LR.4 (see also Eclipse$^{TM}$ 2008).

The dose model is a superposition-convolution algorithm, i.e. a superposition of individual 3D proton beamlets convolved with the fluence at the position of the beamlet. The term 'fluence' refers here to the undisturbed fluence in air $\Phi_{air}$. The calculation and configuration of $\Phi_{air}$ is described in detail by Schaffner (2008). The fluence in air incorporates all effects, which contribute to the lateral distribution of the protons when they exit the beamline. The main effects modeled in the fluence are the initial lateral penumbra (following the effective source size concept introduced by Hong et al 1996), the scatter in the compensator and the phase space/weight of scanning pencil beams. The dose calculation concept follows previously published approaches (Schaffner et al 1999, Ulmer et al 2005). In short, it is given by the following formulas:

$$\begin{aligned} D_{beamlet}(r,z) = \frac{1}{\rho_{H_2O}} & \left[ S_{pp}(d(z)) + S_{rp}(d(z)) + S_{sp,n}(d(z)) \right] \cdot K_{lat,prim}(r,d(z)) \\ & + \left[ S_{sp,r}(d(z)) + S_{heavy}(d(z)) \right] \cdot K_{lat,sec}(r,d(z)) \end{aligned} \quad (30)$$

$$D_{total}(x,y,z) = \sum_{i,j} \Phi_{air}(x_i, y_j, z) \cdot D_{beamlet,ij}(r(x,y),z) \quad (31)$$

For the practical computation, we substitute the convolution by a sum over beamlets at each point of the calculation grid. $\Phi_{air}$ is always taken at the position corresponding to the central axis of each beamlet, i.e. $x_i$ and $y_j$ for the beamlet ij. Please note that $\Phi_0$ appearing in sections 2.2 – 2.6 is dimensionless and has to be put *1*. To save computation time, we describe the lateral extension of recoil protons by the same kernel as for primary protons, since it can be assumed, that their production follows the distribution of primary protons and the energy is deposited locally. Heavy recoils deposit most of their dose through the $\beta^+$-decay and annihilation or by neutron emission (see listing (27)). This means that their lateral distribution is very broad. Due to the very small overall contribution of heavy recoils, we simply add the dose of heavy recoils to the dose deposited by secondary reaction protons, which have in general the broadest distribution. Another option would be to use a weighted sum of the squared contributions from Eq. (31), which is done in Fig. 5. The 3D beamlet (Eq. (30)) is calculated from the results in the previous sections and LR.4 using the simplifications discussed above.

The distance from the central axis of the beamlet is denoted by r, z is the position along the central axis of the beamlet, and d(z) is the water-equivalent-distance up to the position z. Fig. 5 plots the building blocks of the beamlet calculation; the stopping power distribution in depth and the lateral distribution due to scattering in water. $\rho_{H2O}$ (in Eq. (30)) is the density in water, since we are calculating dose to water and not dose to medium – thus following the convention used in photon dose calculation algorithms. The path length correction or scaling of the beamlet in depth for non-water media is taken into account by using d(z) instead of a distance from the surface in the beamlet calculation. The path length correction is applied in the same way for the scaling in depth and in the lateral distribution. An improved model for the scaling of the lateral distribution in heterogeneous media has been presented by Szymanowski et al (2002).

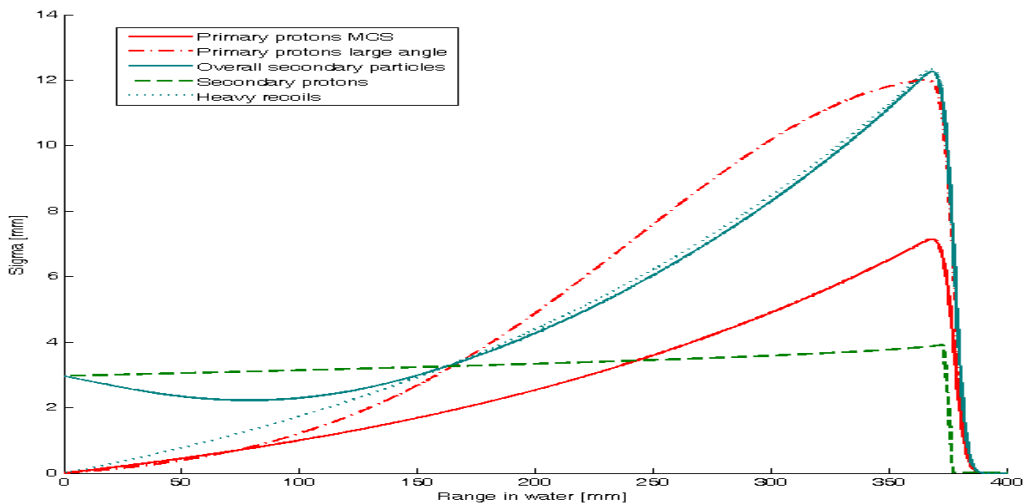

Fig. 5. Example for the lateral dose distribution in a beamlet for a 250 MeV proton beam. The beam width σ is equal to $\tau/\sqrt{2}$. Secondary protons exhibit a non-zero width at the entrance. This is due to the contribution of $\tau_{heavy}$, see Eq. (L30). The emission of heavy recoil particles from a nuclear reaction is assumed to be isotropic.

**3. Results**

*3.1  Fitting of the depth dose model to measured pristine Bragg peaks*

We used measured pristine Bragg peaks to test our model. The depth dose measurements were performed differently, depending on the delivery technique of the treatment machine. Double scattering pristine Bragg peak curves are measured with a thimble ionization chamber in a broad beam. In order to compare them to our model, we correct the measured pristine Bragg peak curve for SAD effects by using the $1/r^2$–law and shift the measurement depth by the water-equivalent thickness of the nozzle (NeT). The values of the NeT was provided by the machine manufacturer and always lead to a good agreement (better than 1 MeV) between the energy obtained from our fits and the nominal energy as provided by the manufacturers. In the case of uniform and modulated scanning beams, the pristine Bragg peak is typically measured from a single static pencil beam with a large parallel plate chamber. Only the shift in depth with the water-equivalent thickness of the nozzle applies here. When fitting the beamlet model to a measured depth dose curve, we only allow a variation of following parameters:

- Spectrum of the initial beam $\tau_{in}$ – the main impact of the beamline properties on the shape of the depth dose curve (see also Fig. 11).

- The normalization factor $N_{abs}$, which allows the conversion of the calculated depth dose from MeV/cm to the measured unit of Gy/MU.

- Energy of the initial beam. The energy is fitted, and usually agrees to better than 1 MeV with the energy claimed by the machine manufacturer. Start values of energy fitting are either specifications of the manufacturer or the expected $R_{csda}$ (the 90 % value behind the Bragg peak).

These three parameters have the most impact on the shape and the absolute scaling of the depth-dose curve. *Minor* corrections are possible by allowing the following parameters to vary:

- Fraction of secondary protons reaching the water phantom expressed by the factor $\upsilon$. A possible option (1) is to assume, that a certain percentage of secondary protons are lost along the beamline due to the fact that they are scattered broader than the primary component and, therefore, hit the primary collimator. The fraction of secondary protons is only fixed to 1 for scanning beamlines (as in Figures 6 – 7). A further option (2) is the application of a specific Monte-Carlo code (Matsinos 2008) to simulate the beamline characteristics (if all necessary data are available) and to determine to the associated phase space. The latter method provides a sufficient calculation basis, whereas the first way is a pure fitting by 'try and error'.

- Landau parameter ($C_{Lan1}$): There is a comparably small dependence of the amount of Landau correction on the beamline. However, compared to $\tau_{in}$ its impact is small (see also Figures 12 - 13).

The depth-dose model is fitted in a two-step approach to the measured pristine Bragg curves after their processing as explained above. In a first round of fitting, the free parameters are the energy at nozzle entrance ($E_0$), the initial range spectrum $\tau_{in}$ and a normalization factor. As already mentioned, we allow fitting the contribution of secondary protons by a variation of $\upsilon$ according to option (1), which enters the fluence of secondary protons $\Phi_{sp,r}$ (see e.g. equations (L8 – L9)). The use of this option makes sense, whenever a considerable fraction of secondary protons may not been detected, as they might have been stopped along the beamline and option (2) cannot be applied. This condition typically applies to scattering beamlines. In the second round of fitting, we can fit $C_{Lan1}$ freely while allowing only a 2% variation of the energy, the range spectrum and the normalization, and a 10% variation of the contribution of secondary protons by the user. Some results of the fitting of our model to measured pristine Bragg curves are shown in Figure 6 (modulated scanning, high-energy beam from an Accel machine), Figure 7 (modulated scanning, low-energy beam from an Accel machine) and Figure 8 (double scattering, high-energy beam from an IBA machine). The match between calculation and model is excellent for all cases. It has to be pointed out, that the term $I_{Lan1}$ depends on the residual range ($R_{CSDA}$ in Eq. (28)) of the beam when entering the medium. In scattering techniques, there is always a considerable amount of absorption through the elements of the nozzle. This means that $R_{CSDA}$ must be replaced by $R_{CSDA}(E_0) - NeT$. This shift in the position of the Landau correction can be seen in Fig. 8, where the zero position of the depth axis corresponds to the entry position of the beam into the nozzle. Measurements and calculation only start after the range shift through the nozzle elements.

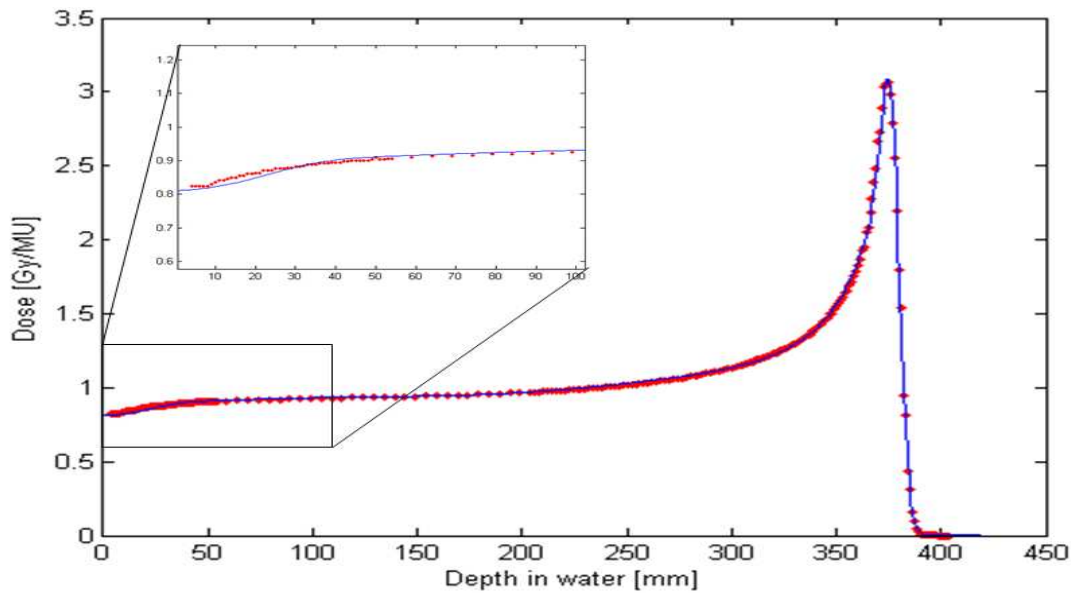

Fig. 6. Comparison between measured and calculated pristine Bragg peak. The dose is measured with a large parallel plate chamber across a single 250 MeV pencil beam delivered by the Accel machine. The buildup is most visible for high-energy beams without absorbing material in the beamline; it is well described by the model.

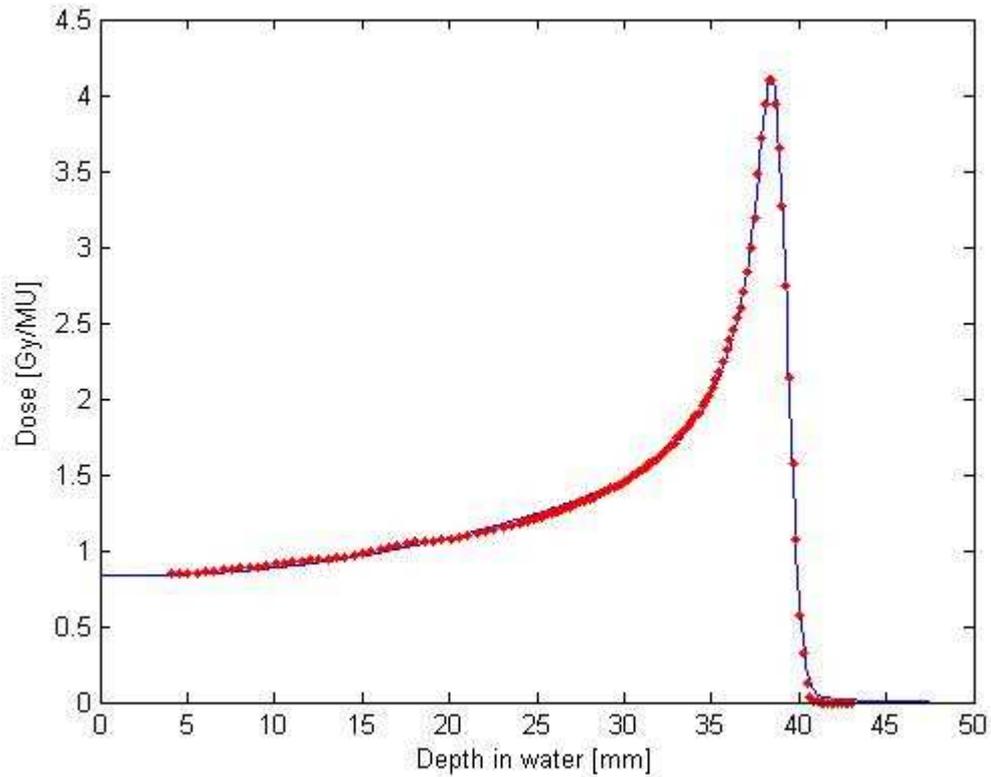

Fig. 7. Comparison of measured and calculated pristine Bragg peak. The dose is measured with a large parallel plate chamber across a single 68 MeV pencil beam delivered by the Accel machine.

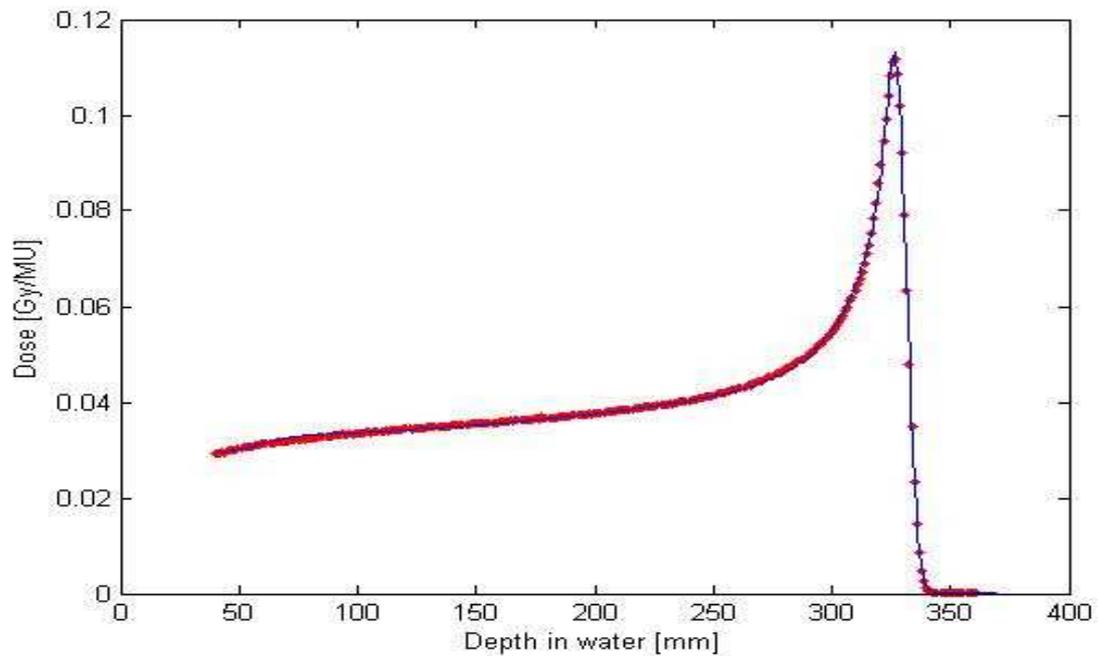

Fig. 8. Comparison between a calculated and measured pristine Bragg peak for the IBA double scattering beamline at the Korean National Cancer Center (KNCC). The initial beam energy is 230 MeV. The measured data points are corrected for SAD effects and shifted in depth by the nozzle-equivalent thickness of 40.7 mm.

Note that the buildup effect is also clearly visible in this scattering beamline and should not be neglected for higher energies. The field-size (10 x 10 cm$^2$) did not affect this Bragg curve. However, field-sizes smaller than the present one can also affect the initial plateau (Ulmer and Schaffner 2006).

Since the adaptation of the calculation model to measured data is an essential technical aspect of therapy planning, we refer to the Eclipse manual (Eclipse$^{TM}$ 2008) for further information.

*3.2   Lateral scattering*

The experimental verification of the lateral scattering calculation is very challenging due to the small contributions of the large angle scattered primary protons and the secondary protons. Furthermore, there are other contributions to the lateral distribution of protons (initial phase space of a scanned beam or effective source size and block scattering in a scattered beam) which affect the measurement. The following plots show a comparison between calculated beam width and beam width fit to the measurements using our theory.

Figure 9 shows how small the contributions of larger order scattering are. These contributions may be negligible in the situation measured here, but play a more important role in higher energies and also if the scatter happens in a range shifter at some distance from the patient. Especially in the latter case, it will affect MU calculations (Pedroni et al 2005). This figure represents one characteristic example of a series of measurements performed with a portal vision imaging system (Varian PVI$^{TM}$) at a scanning beam accelerator (Accel) at various depths and initial energies. Although detailed presentation goes beyond the scope of this study, we note that the initial phase space of the beamlets can be accounted for by the initial fluence map $\Phi_{air}$ referring to the number of protons and to $\tau_{Range}$ (Eq. 17)). As pointed out in LR.4 the lateral scattering represented by kernels $K_{lat,prim}$ and $K_{lat,sec}$ in Eq. (30) and the rms-values $\tau_{LAT}$,…, etc have to be modified. In the domain from the Bragg peak to the distal end these values have to be subjected to a convolution to include energy/range straggling and phase space properties (equations L36 – L41), which provide a cigar-shaped profile of a beamlet. These aspects are preferably of technical nature of the implementation (see Eclipse$^{TM}$ 2008 and Schaffner 2008).

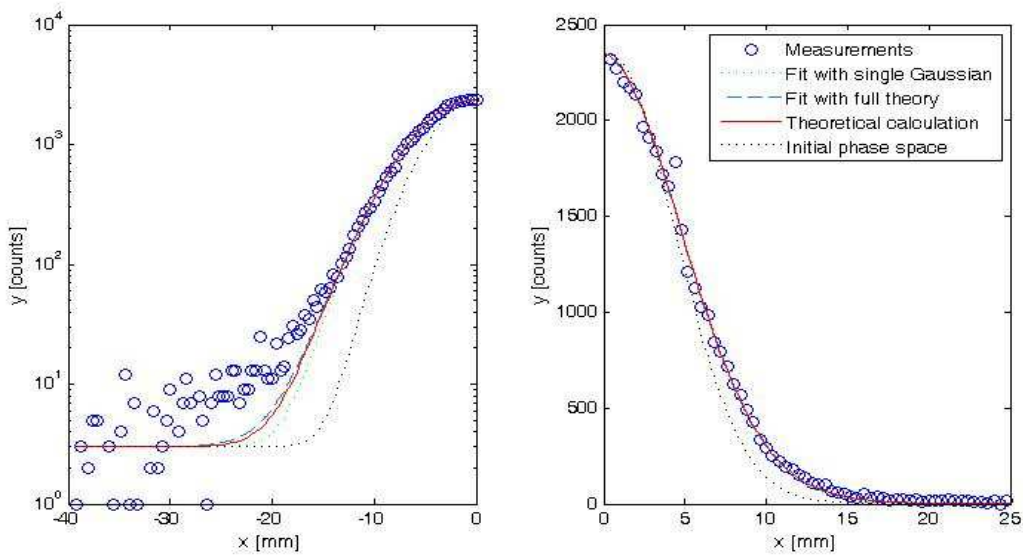

Fig. 9. Logarithmic and linear plots of a spot-size measurement of a 160 MeV beam at a water-equivalent depth of 151 mm for an Accel machine. The black dotted line shows the initial phase space, which is subtracted quadratically from the other contributions.

A very good test of modeling of lateral scattering in a calculation model is the depth dose curve of a very small proton beam, i.e., the field-size is of the order of the rms values $\tau_{Lat}$ of the scatter kernel (see LR.4) . Fig. 10 presents such an example; the agreement with measurement data is excellent. It is only necessary that the air gap (distance between collimator and impinging surface) is sufficiently long to prevent disturbances of scatter protons produced by the collimator wall.

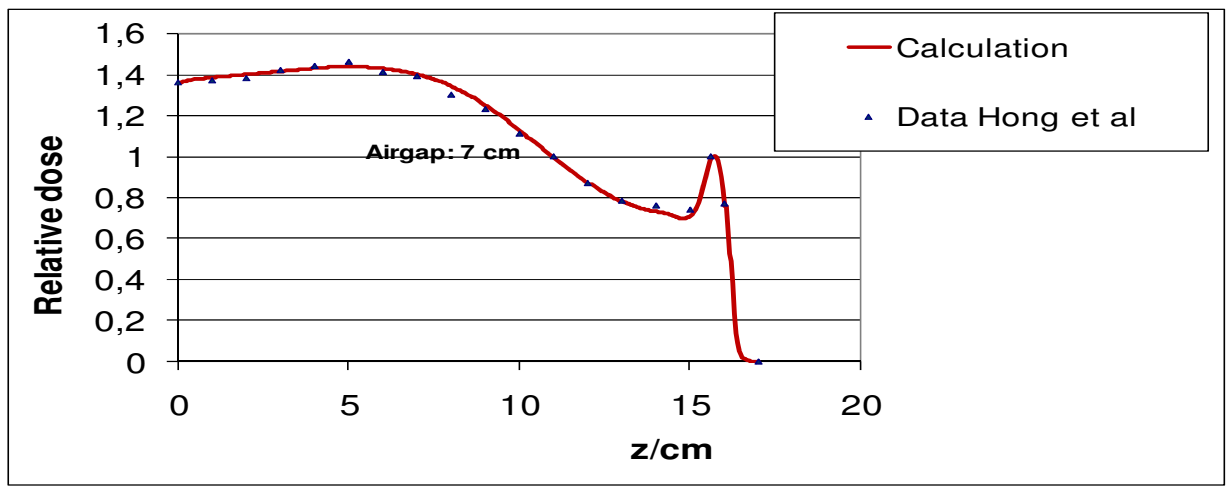

Fig.10. 158 MeV proton beam passing through a cylindrical collimator with radius r = 2.4 mm. Measurement data have been obtained by the Harvard cyclotron.

**4. Discussion**

Looking at the results for the fitted parameter, we typically find agreement between our fitted energy and the energy stated by the machine manufacturer to better than 1 MeV.

Interesting observations can be made by plotting the initial range spectrum $\sigma_{in} = \tau_{in}/\sqrt{2}$ as a function of the nominal energy for a number of different machines and techniques (Figure 11). The Hitachi machine is a synchrotron, which produces naturally a narrow Bragg peak. The other machines are cyclotrons. The lower energies are obtained by degrading the beam from the cyclotron. This process creates a broad energy spectrum, which needs to be narrowed through energy selection slits. The setting of the energy selection slits is a compromise between the width of the Bragg peak and the beam intensity. It seems that this compromise is found at a very similar width for the three machines by Accel, PSI and the IBA machines. The width of the PSI beam is somewhat broader, probably due to the fact that the data from the PSI beamline originate from a 600 MeV physics research accelerator and therefore was degraded much more than the beams from the other machines.

The data points from the Hitachi machine show that the range spectrum seems to increase with the amount of high density material in the beamline (i.e. the large field configuration ($\square$) has more lead than the medium field size ($\diamond$); the scanning beamline ($\nabla$) has no extra material). This shows nicely, that range straggling in higher density materials is larger than in the same (water-equivalent) amount of low density materials. Since our model does not distinguish between different compositions of the beamline (i.e. $\tau_{straggle}$ from Eq. (7) is not changed), the additional straggling component is taken into account by the fitting procedure in beam configuration through an increase in $\tau_{in}$.

We found that the resulting range spectra as a function of the nominal energy can be fitted well by a 3$^{rd}$ degree polynomial for both, synchrotron and cyclotron. This fact allows us to easily model the pristine Bragg peak curves for intermediate (not configured) energies.

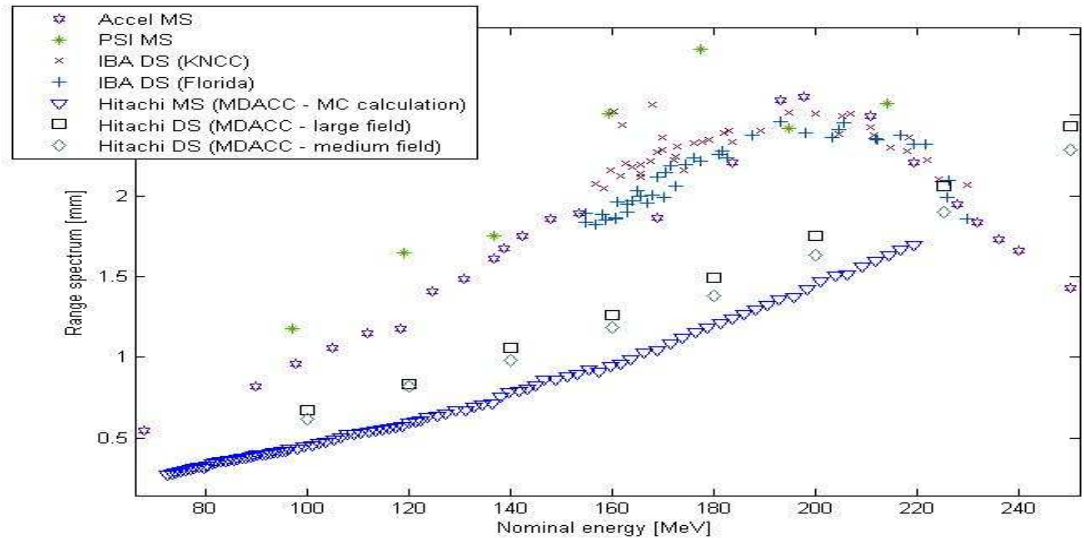

Fig. 11. Range spectra ($\tau_{in}$) obtained from our depth-dose model for a variety of treatment machines and techniques (MS: Modulated scanning, DS: Doubles scattering). It is well known, that a synchrotron (Hitachi) usually creates a smaller initial range spectrum, than a cyclotron (all other machines). It is interesting to observe that all cyclotrons produce similar range spectra. Note: The Hitachi MS data is obtained from Monte Carlos simulated pristine Bragg peak curves; measured data has not yet been available to us.

The fitting of the Landau parameter $C_{Lan1}$ for double scattering (Figure 12) shows quite a bit of scatter between different machines and also for different configurations of the beamline (often called 'options') within the same machine. However, the magnitude of $C_{Lan1}$ is the same for all machines and there is a trend towards a minimum at 100-150 mm residual range. The effect of these variations of $C_{Lan1}$ on the total depth dose calculation is very small and we normally use the Nominal $C_{Lan1}$ according Eq. (28) for scattering and uniform scanning dose calculations.

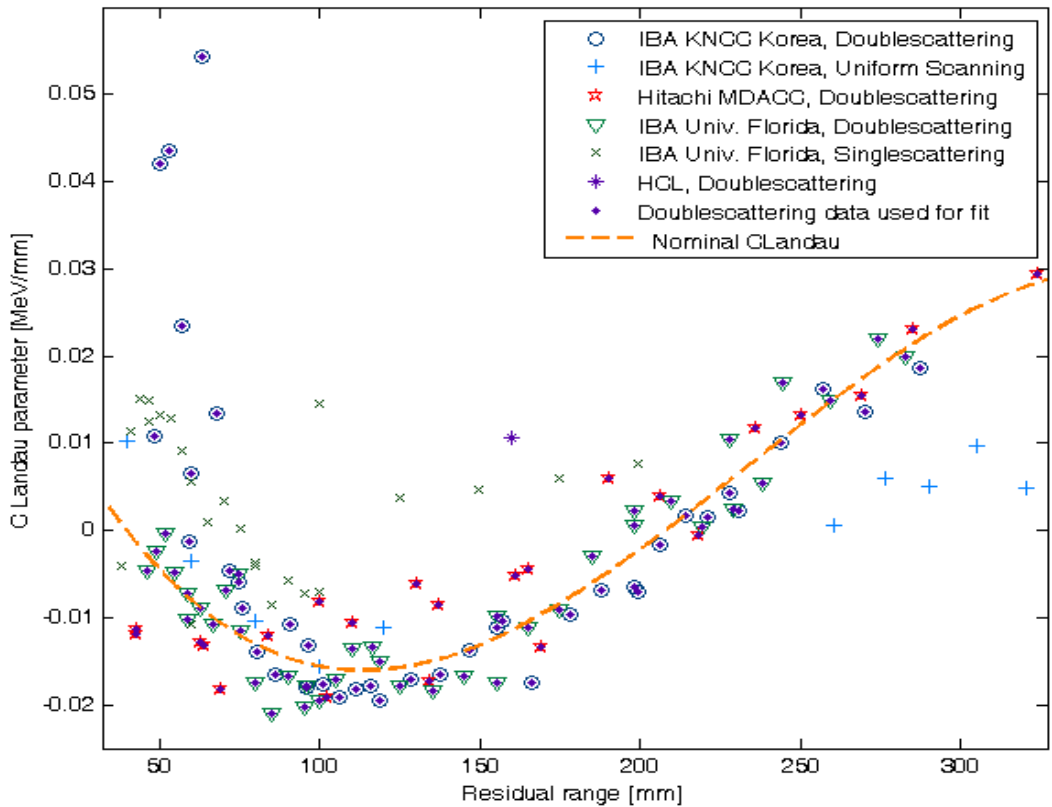

Fig. 12. Fit results for $C_{Lan1}$ for a number of different double-scattering beamlines and for one uniform scanning beamline. The line corresponds to $C_{Lan1}$ of Eq. (28). It has been obtained from a polynomial fit to the data sets Hitachi MDACC, IBA KNCC (double scattering) and IBA Florida. The other data sets are plotted for comparison only. The pristine Bragg peak measurements of some other double scattering beamlines could not be measured up to the surface of the water phantom due to a restriction to horizontal beam geometry. This means, that the most relevant data points for the fitting of $C_{Lan1}$ were missing. Note that the fitting procedure of $C_{Lan1}$ is implemented due to possible tails of the beamline modulating the energy spectrum. Only for monoenergetic protons $C_{Lan1}$ can be restricted to theoretical values.

The results for the pristine Bragg peak measurements from the modulated scanning beamlines show a much clearer trend for $C_{Lan1}$ as a function of the residual range (Fig. 13). We usually substitute $C_{Lan1}$ from Eq. (28) by another 3$^{rd}$ degree polynomial with parameters obtained from fits to the data points of each machine – as indicated by the solid lines in Fig. 13. In a recent paper, Hollmark at al (2004) point out that, in the domain of the Bragg peak, the Gaussian solution (one Gaussian) is sufficient for both longitudinal energy straggling and lateral scatter. The corresponding arguments are based on the transition of the more general Boltzmann transport theory to the Fermi-Eyges theory in the low-energy limit. However, it appears that the conclusion is only partially true, since the history of the proton track has an influence on the behavior in the low-energy domain (see above results referring to the Landau tails of energy transfer), and, according to results of Matsuura et al (2006), all types of transport equations also have more general solutions than given by one Gaussian in the diffusion limit. Yet the linear combination of two Gaussians with different half-widths, as used in the present study, is not a solution of Fermi-Eyges theory, but a corresponding one of a nonlocal Boltzmann equation. This is an integro-differential equation with different transition probabilities for the local and nonlocal part (long-range interaction). In the diffusion limit, the nonlocal part provides, at least, one additional Gaussian.

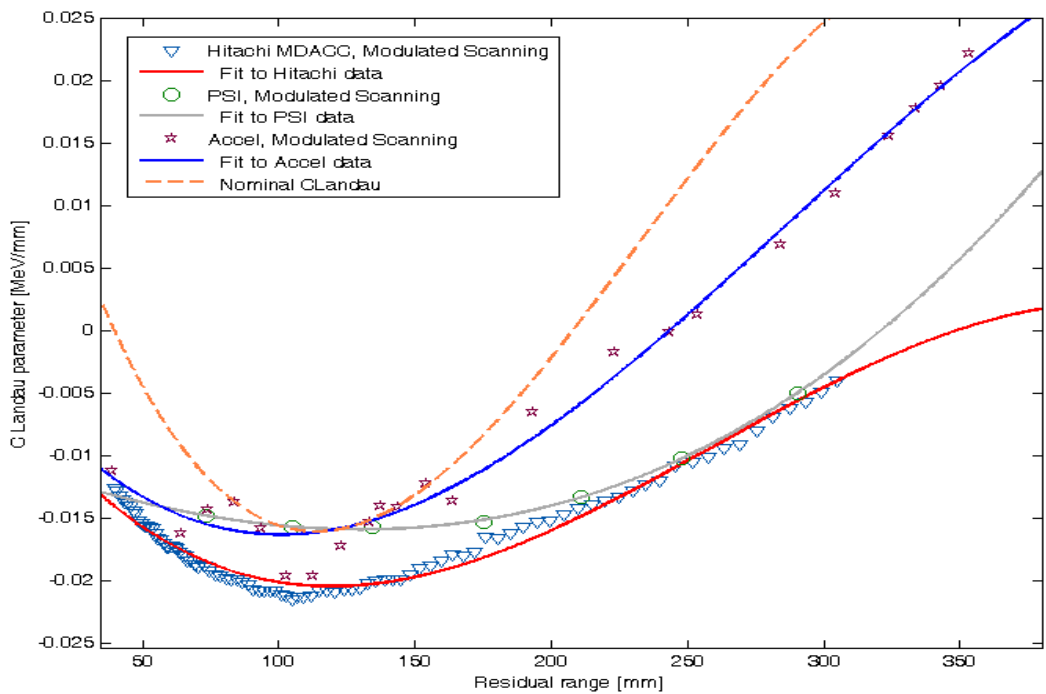

Fig. 13. The value of $C_{Lan1}$ for different modulated-scanning beamlines (Note: Hitachi MDACC originates from Monte Carlo calculated pristine Bragg peak). The fitted lines are $3^{rd}$ degree polynomials. They replace the nominal $C_{Lan1}$ from Eq. (28) in the dose calculation for the respective machines.

## 5. Conclusion

We have developed an analytical model for the depth-dose distribution of a proton beam – the pristine Bragg peak. The model depends on a few beamline-specific parameters (nominal energy, energy/range spread, Landau parameter, contribution of secondary protons), which need to be obtained by fitting the model to the measured pristine Bragg curves. We have shown that the model can reproduce the pristine Bragg curves for different accelerator and beamline designs. An interpolation of the key parameters allows us to accurately calculate any intermediate pristine Bragg peaks; this is particularly important for delivery machines which feature an analog energy tuning.

The lateral distribution of the protons is modeled separately for primary and secondary protons; in order to describe better the large-angle scattering, the lateral distribution of primary protons is modeled by a sum of two Gaussians. However, it has been shown by Pedroni et al (2005) and Kusano et al (2007) that a correct modeling of the large-angle scattered primary protons and the scattering of the secondary protons has an important impact on the determination of the MU factor.

There are two important features, which have to be taken into consideration in an elaborated therapy planning system such as Eclipse$^{TM}$ (2008) and have only been treated here marginally: 1. Heterogeneity problem; 2. Inclusion of collimator/block scatter in broad beams.

*To point 1:* The formulas for E(z), dE(z)/dz, etc. for media different from water have been stated ((equations (1, L4, L5, and table L3)). They can also be handled by the stepwise manner. The difficulty arises for voxel information of CT with regard to the fixation of $E_I$, Z and $A_N$, since CT only provides the electron density. In Eclipse$^{TM}$ a proposal of Schneider et al (1996) is implemented to handle the heterogeneity, and the lateral scattering treatment in the case of heterogeneity is based on the findings of Gottschalk et al (1993), see e.g. Figure L11 in LR.4.

*To point 2:* Collimator/block scatter is also implemented in Eclipse$^{TM}$. The algorithm incorporates a *second source* and is based on studies of Matsinos (2008). The scatter contribution cannot be neglected, since it is responsible for 'horns' in the transverse profiles of broad beams. As already pointed out collimator scatter may also lead to peculiar Bragg curves of very small field-sizes (Ulmer and Schaffner 2006).

**Acknowledgements**

Many thanks go to the following institutions and persons for providing the data used in this paper: IBA (N. Denef, M. Closset, V. Breev and G. Mathot), KNCC (S. B. Lee, D. Shin, J. Kwak, D. Kim), MDACC (R. Zhu, J. Lii, N. Sahoo and clinical physics team, M. Bues and George C. (MC data)), MPRI (W. C. Hsi), PSI Switzerland (T. Lomax), University of Florida (D. Yeung, R. Slopsema), Varian-Accel (J. Heese), Varian iLab (D. Morf). Another 'Thank you' goes to the colleagues B. Schaffner, C. Ritzmann and W. Kaissl for implementations and useful discussions and to the Reviewer of this journal for helpful comments.